# Craters and ring complexes of the North-East Sudanese country

**Amelia Carolina Sparavigna**
Dipartimento di Fisica, Politecnico di Torino
C.so Duca degli Abruzzi 24, Torino, Italy

**Abstract**
We propose a survey of a rocky region in the north-east part of Sudan, using the satellite imagery from Google Maps. In particular we analyse the region which lies to the north of Nakasib Suture. Images reveal craters and ring complexes. To enhance their features, images are processed with a method based on fractional calculus. Two of these structures are proposed as possible impact craters.

The north-east part of Sudan is now claiming a great deal of activity in geological researches. In fact, the Sudanese government engaged a company to explore a rocky area that might hold gas or crude oil [1]. As reported [1], the region has a sedimentary base which follows the same system as the bases in the southern Sudan, probably generating hydrocarbons too. North-east of Sudan is also interesting for mining gold [2]. In fact, the geo-chronological structure of the region is well-known [3-4]. Moreover, the rocky part of this area, the region of the Red Sea Hills, was recently studied to determine the structure of dykes and their effects on groundwater flow [5-6].

In this paper we survey the region which is north of Nakasib Suture, to evidence possible interesting features, as previously done for the Bayuda desert [7]. The Nakasib Suture is a structural belt in the central Red Sea Hills of the Sudan. It represents one of the sutures along which the microplates of the Arabian-Nubian Shield are welded together [8,9].

Among the many interesting landforms that we can see, there are crater-like objects, ring-complexes and large flat circular basins. The survey can be easily performed with the satellite imagery of Google Maps or ACME Mapper, powered by Google. Images are the same for both services and can reach very high resolutions. For the analysis of landforms, ACME Mapper seems to be more easy and quick. Since land domains can have quite different size, the zoom level necessary to observe the structure is usually different for each domain.

When a region shows interesting features, the quality of the mage at a proper resolution level can be enhanced using several image processing methods. In this paper we use a tool based on the calculus of the fractional gradient of the colour tone map, to enhance the edges of domains without damaging the overall visibility of images. This tool was prepared to be freely used under Windows .NET, and since it was firstly developed for astronomy, it was named AstroFracTool. For the mages shown in this paper, the fractional and visibility parameters were chosen as $v=0.5$, $\alpha=0.5$ (see Ref.10 for a detailed discussion). After this processing with AstroFracTool, the resulting image is subjected to an adjustment of contrast and brightness with GIMP software.

Before starting the description of some craters evidenced by the survey, let us spend few words on their shape. Craters can be created by collisions of small extraterrestrial bodies with the Earth, or built by volcanic activity. As an impact crater, the corresponding form generally has a roughly circular outline and a raised rim. The size can be very large for multi-ringed basins. The shape is generally circular, although elliptical impact craters are known due to oblique collisions. In addition, impact craters can look no longer circular because of tectonic deformations [11]. A well-preserved simple impact crater has a bowl-shaped form, usually less than 2 km in diameter, with relatively high depth to diameter ratio. Complex craters have a low depth to diameter ratio and possess central uplifts. Multi-ring craters or basins have depth to diameter ratios like those of complex impact craters, but they possess concentric rings, marked by normal faults with downward motion toward crater center [11]. A satellite survey is then the best method to locate craters and catalogue them, and, free for everybody connected to the Web [12].

A volcanic crater is a roughly circular, rimmed structure usually located at the summit of the volcanic cone built by erupted materials. Craters can also form when volcano collapses. Another kind of volcanic crater is a maar crater, formed during phreato-magmatic eruptions, from explosion caused by groundwaters coming into contact with the magma. The resulting blow-out causes a circular depression to form. A volcanic crater then can be of large dimensions, sometimes of great depth. Maars are shallow, flat-floored, usually full of water. The north-east of Sudan displays many circular large basins, which, from satellite imagery appear as full of sediments, as maar craters without water.

Both impact and volcanic craters are subjected to weathering and tectonic actions. It turns out that crater-like landforms observed by satellites can have a quite different origin. The local search for impact evidences, such as meteoritic fragments and shocked and shock-melted materials, is therefore fundamental . In any case, as suggested by D. Cohen, University of New South Wales, in [13], a database of satellite imagery could be useful for comparing the different impact and volcanic structures.

Let us discuss two structures in north-east of Sudan which have a crater-like form. Figure 1 shows the region near the Nakasib Suture containing one of them: the figure is obtained processing as previously discussed an image from Google Maps. The box exhibits the structure of interest, near the Jabal Ababad, which seems a ring or doughnut, of approximately 3 km diameter. Figure 2 shows the marked region with increasing resolution. In the upper left panel there is the satellite image from Google, whereas in the upper right panel, the terrain as displayed by the Google tool . The coordinates of this crater are approximately 19.213 N, 35.983 E. As suggested in [13], the comparison with the surrounding land texture helps to conclude that the domain in Fig.1 and 2 is unique in the considered region.

Another crater-like structure, not so well-defined as that previously shown, is located approximately at 21.290 N, 33.992 E. Figure 3 reproduces the region as appears after image processing. Zooming with Google Mapper on the crater, we have the image on the left of Figure 4. The terrain is shown in the middle. In the right panel, the image after processing. The diameter is large, 6 or 7 km. The comparison with the surrounding land texture shows that this is a circular domain, quite different from the background. This is the same conclusion we had for the object in Fig.1 and 2: these crater-like forms seem to be true after comparing with the local texture.

Let us conclude our survey discussing Figure 5, showing some of the several ring complexes that one can observe in the region. The sizes are ranging from 2 to 5 km. These ring complexes are examples of forms created from an igneous intrusion, producing ring dikes and cone sheets. The texture of these complexes recall that of the toric domains observed with the polarized microscope in liquid crystals, and created by the growth, that is an intrusion, of a smectic layered phase in the isotropic melted phase of the liquid crystal material [14-16].

For what concerns the two crater-like forms in Figs.1-4, the author cannot tell whether these structures have been already reported in literature on impact craters or not. This literature is huge: in the database [17,18], they are not reported. Let the author suggest these structures as possible impact craters, and, at the same time, remark that the aim of the paper, besides the location of peculiar domains, is that of showing a possible use of public maps to investigate and compare the Earth landforms, after increasing the resolution with free processing tools.

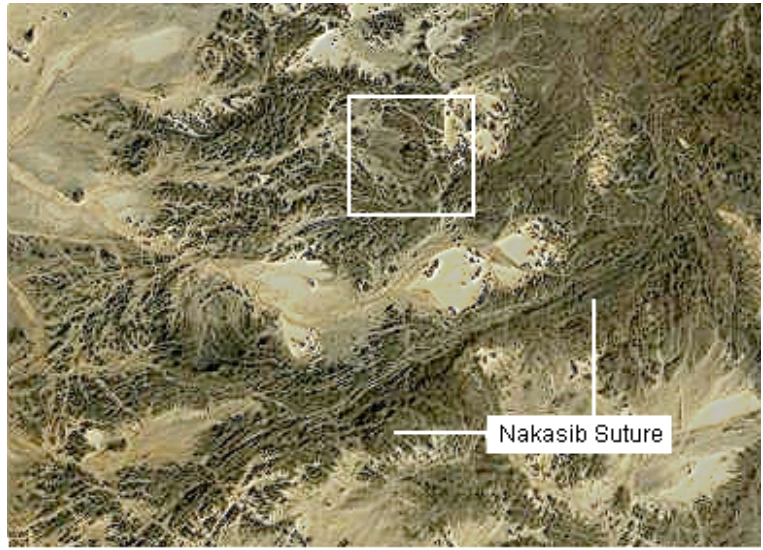

Fig.1. The figure shows the region near the Nakasib Suture. The image is obtained after processing a satellite image from Google Maps. The box is pointing out an interesting ring structure, near the Jabal Ababad. This ring or doughnut has a diameter of approximately 3 km.

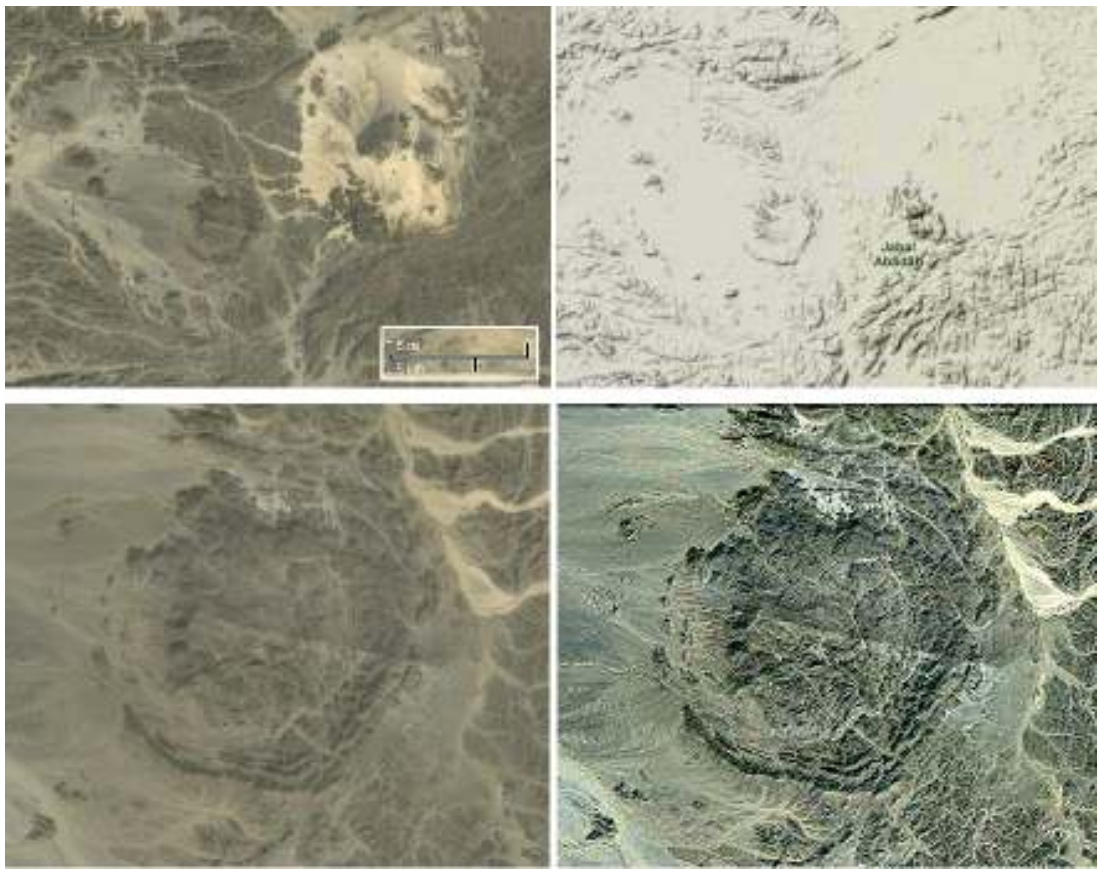

Fig.2. The figure shows the marked region of Fig.1 with increasing resolution. In left panels, the satellite images from Google. In the upper right panel, the terrain as obtained from the same search tool. The position of the crater is approximately 19.213 N, 35.983 E. The lower right panel shows the image as obtained after processing (the diametral line is a defect of the original image).

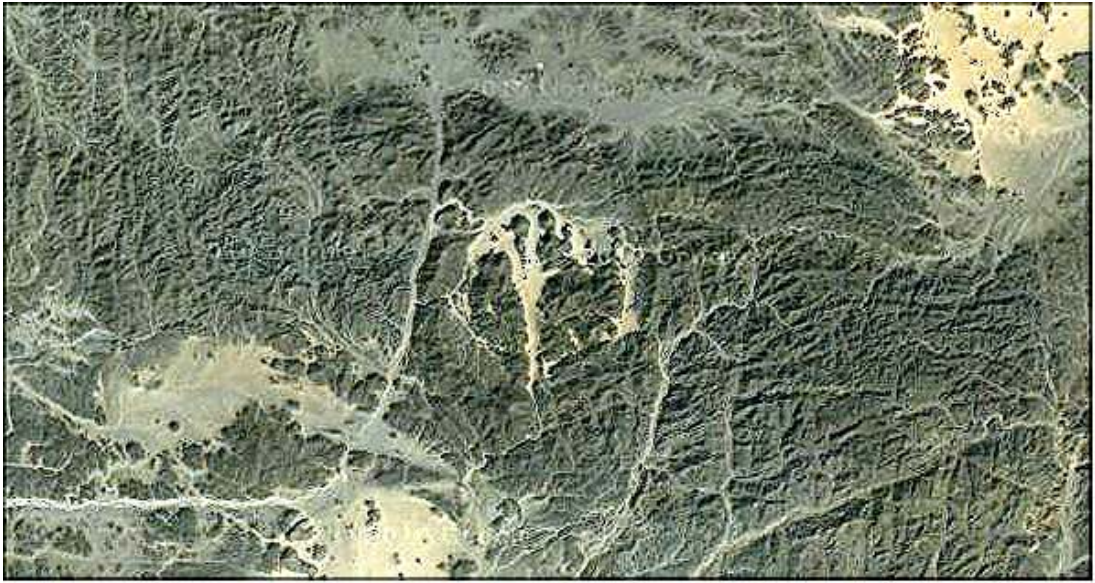

Fig.3. Another crater-like structure, not so well-defined as that shown in Fig.1 and2, located approximately at 21.290 N, 33.992 E. The image shows the region as appears after image processing. This circular crater is embedded in a land, the texture of which is not displaying circular features.

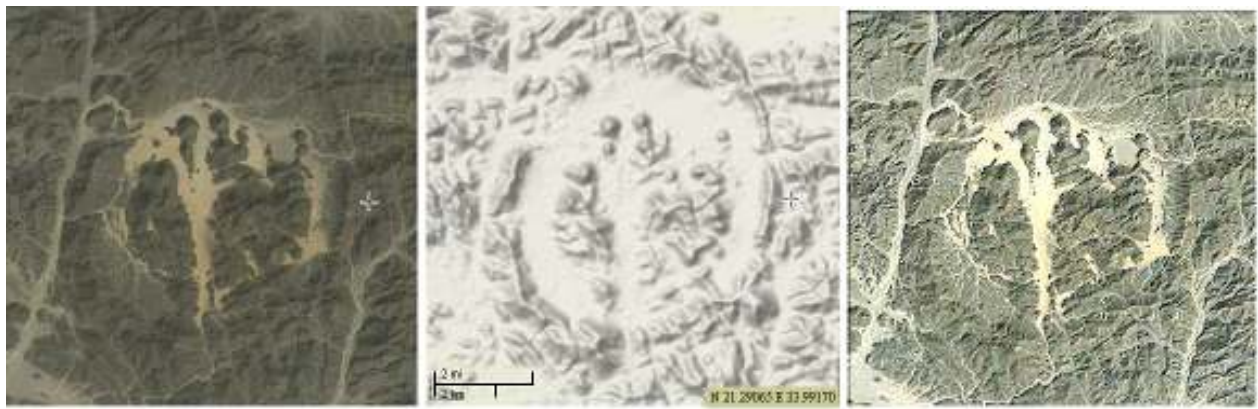

Fig.4. Zooming on the crater with Google Maps, we see the image on the left. The terrain obtained by means of the Google tool is shown in the middle. In the right panel, the image after processing. The diameter is large, 6 or 7 km. Coordinates are 21.290 N, 33.992 E.

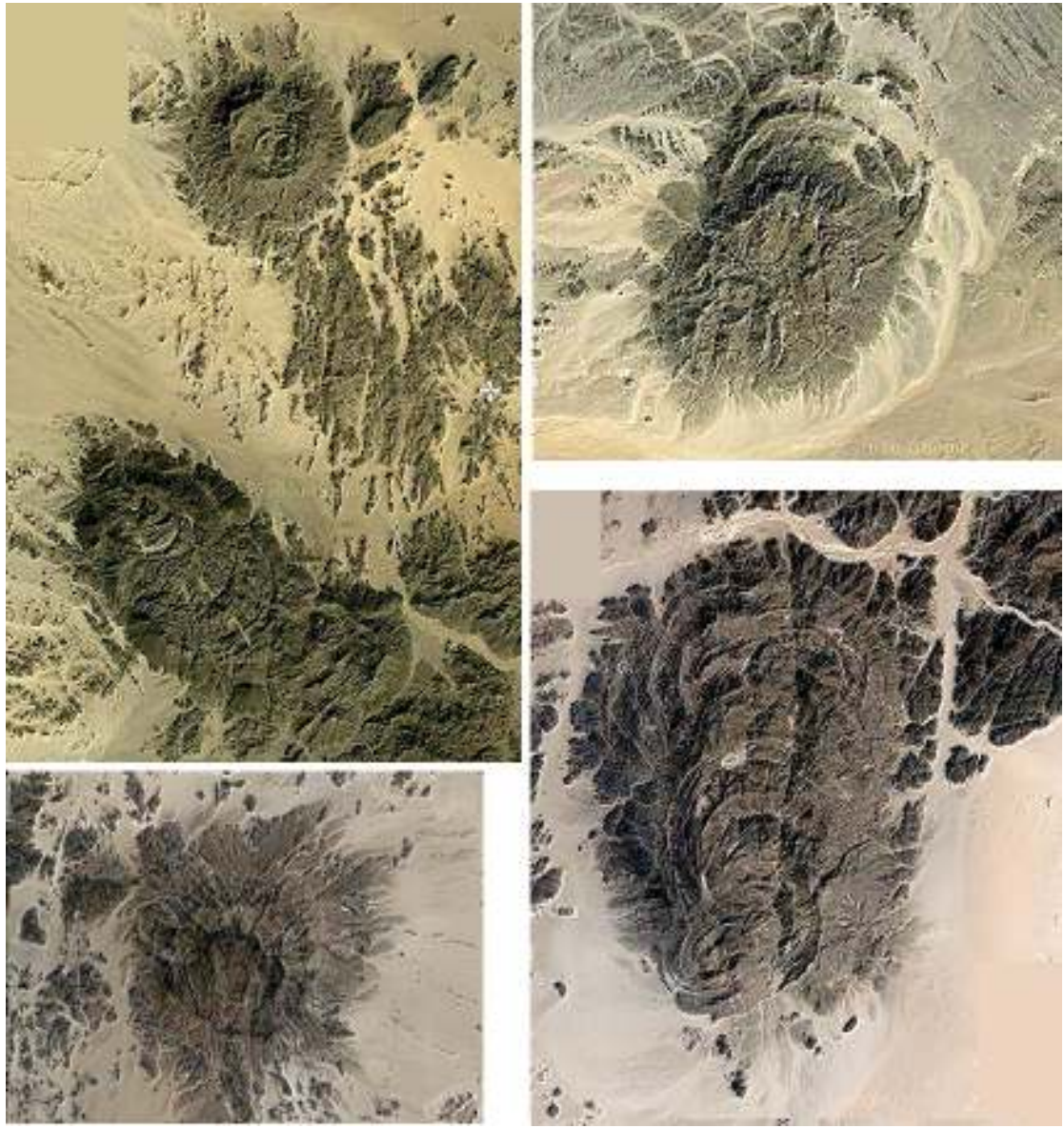

Fig.5. Images show some of the several ring complexes that one can observe in the north-east Sudanese region. The sizes are ranging from 2 to 5 km.